\documentclass[showpacs,twocolumn,aps,preprintnumbers,letterpaper]{revtex4}
\usepackage{amsmath,amssymb}
\usepackage{epsfig}
\usepackage{graphicx}
\usepackage{tabularx}
\usepackage{amsmath}
\usepackage{slashed}
\usepackage{amsfonts}
\usepackage{epstopdf}
\usepackage{color}
\usepackage[dvipsnames]{xcolor}
\usepackage[caption=false]{subfig}
\usepackage[pdftex, pdfborder= 0 0 0, citecolor=blue, urlcolor=blue, linkcolor=blue, colorlinks=true, bookmarksopen=true]{hyperref}

\newcommand{\be}{\begin{equation}}
\newcommand{\ee}{\end{equation}}
\newcommand{\bear}{\begin{eqnarray}}
\newcommand{\eear}{\end{eqnarray}}
\newcommand{\ba}{\begin{array}}
\newcommand{\ea}{\end{array}}

\def\be{\begin{eqnarray}}
\def\ee{\end{eqnarray}}
\def\bea{\be}
\def\eea{\ee}

\def\roughly#1{\mathrel{\raise.3ex\hbox{$#1$\kern-.75em%
\lower1ex\hbox{$\sim$}}}}

\def\abs#1{{\left| #1 \right|}}

  \long\def\comment#1{ }

  \newcommand{\Tr}{{\rm Tr}}

  \newcommand{\beq}{\begin{eqnarray}}
  \newcommand{\eeq}{\end{eqnarray}}
  
 \def\simge{\mathrel{%
   \rlap{\raise 0.511ex \hbox{$>$}}{\lower 0.511ex \hbox{$\sim$}}}}
\def\simle{\mathrel{
   \rlap{\raise 0.511ex \hbox{$<$}}{\lower 0.511ex \hbox{$\sim$}}}}

\renewcommand{\vec}[1]{{\mathbf{#1}}}

\begin{document}

\title{$J/\psi$ near threshold in holographic QCD:\\
A and D gravitational form factors}

\author{Kiminad A. Mamo}
\email{kmamo@anl.gov}
\affiliation{Physics Division, Argonne National Laboratory, Argonne, Illinois 60439, USA
}

\author{Ismail Zahed}
\email{ismail.zahed@stonybrook.edu}
\affiliation{Center for Nuclear Theory, Department of Physics and Astronomy, Stony Brook University, Stony Brook, New York 11794-3800, USA}



\date{\today}
\begin{abstract}
The diffractive photoproduction of $J/\Psi$ on a nucleon,  is mostly due to  gluonic exchanges at all $\sqrt{s}$.
In holographic QCD (large number of colors and strong $^\prime$t Hooft coupling), these exchanges are captured by  gravitons
near threshold, and their  Reggeized Pomeron form asymptotically. We revisit our holographic analysis 
of the   A and D gravitational form factors in light of the new lattice data, and use them
to refine our predictions for the photoproduction of $J/\Psi$ near threshold, and the comparison to the
GlueX data. We use these results to estimate the scalar and mass radii of the nucleon, and describe the
gravitational pressure and shear across a nucleon.
\end{abstract}


\maketitle

\setcounter{footnote}{0}


\section{Introduction}

The  breaking of conformal and chiral symmetries in QCD are at the origin of  most of the visible mass in the Universe, a remarkable  quantum
feat, starting from a classical field theory  without explicit mass~\cite{Wilczek:2012sb,Roberts:2021xnz} (and references therein).  At low resolution, 
these phenomena are  tied semi-classically  to topologically active gauge fluctuations: eternal tunneling events between gauge vacuua with different 
winding numbers, also known as  instantons and anti-instantons~\cite{Shuryak:2018fjr,Zahed:2021fxk} (and references therein), with strong evidence
from numerical lattice simulations using cooling procedures~\cite{Leinweber:1999cw,Biddle_2020}.

An empirical understanding of these mechanisms can be made using dedicated electron machines~\cite{Hafidi:2017bsg,Ali:2019lzf,Meziani:2020oks,Anderle:2021wcy}. 
Recently, the GlueX collaboration~\cite{Ali:2019lzf} at JLAB has reported threshold data for
photo-production of charmonium $J/\Psi$ that may start to lift the lid on some of these fundamental questions.
Indeed, near threshold the diffractive  production of charmonia and bottomia is dominated by gluons or glueballs, and
currently measurable. 

In a recent analysis of the GlueX data using a holographic construction, we have shown~\cite{Mamo:2019mka} 
that the threshold differential cross section is only sensitive to the tensor gravitational form factor, and suggested
that  this tensor  form factor or A-term  is extractable from the current data under a minimal but universal set of holographic assumptions. This allows for a first extraction
of the tensor mass radius among other things. Remarkably, the holographic construction ties the A- and D-gravitational
form factors, thereby allowing for the extraction of the gluonic pressure and shear inside the proton. Similar ideas were
also explored in a hybrid form of holography in~\cite{Hatta:2018ina}, and in general in~\cite{Kharzeev:2021qkd,Ji:2021mtz,Hatta:2021can,Guo:2021ibg,Sun:2021gmi,Wang:2022vhr}.

In this work, we revisit our holographic analysis of the A and D gravitational form factors~\cite{Mamo:2019mka,Mamo:2021krl}, by improving
our determination of $A(0)$ and $D(0)$ from a comparison to the newly released lattice data~\cite{Pefkou:2021fni}.
We will use these improved  results, to reanalyze the threshold photoproduction of charmonium as reported by the GlueX
collaboration, and the interplay of the A and D form factors in these results. 

In section~\ref{ABC}, we briefly review the holographic arguments for the A and D form factors for the proton, and compare them 
to the newly released lattice results for the gravitational form factors of the proton~\cite{Pefkou:2021fni}.
In section~\ref{EE}, we show that the A and D form factors control the electro-production amplitude of charmonium near threshold,
with the D contribution suppressed by the squared skewness of the process.  We use these form factors to characterize the 
scalar and mass radii, and the gravitational pressure and shear across a nucleon.
We reduce the electroproduction process to the photon point, and revisit our comparison to the GlueX data. Our conclusions are in section~\ref{CC}.
Some of the details regarding parts of the  derivations are Appended.

\section{Gravitational form factors}~\label{ABC}

The standard decomposition of the  energy-momentum form factor in a nucleon state is~\cite{Kobzarev:1962wt,Pagels:1966zza,Carruthers:1971uy,Polyakov:2018zvc}

\begin{widetext}
\bea
\label{A1}
\left<p_2|T^{\mu\nu}(0)|p_1\right>=\overline{u}(p_2)\left(
A(k)\gamma^{(\mu}p^{\nu)}+B(k)\frac{ip^{(\mu}\sigma^{\nu)\alpha}k_\alpha}{2m_N}
+C(k)\frac{k^\mu k^\nu-\eta^{\mu\nu}k^2}{m_N}\right)u(p_1)\,,
\eea
\end{widetext}
with  $a^{(\mu}b^{\nu)}=  \frac 12 (a^\mu b^\nu+a^\nu b^\mu)$, 
 $k^2=(p_2-p_1)^2=t$, $p=(p_1+p_2)/2$  and the normalization $\overline u u=2m_N$. 
(\ref{A1}) is conserved and tracefull. Throughout, $D(k)=4C(k)$ will be used interchangeably.
In holography,  (\ref{A1}) sources the metric fluctuations in bulk, 
$g_{MN}(z)\rightarrow g_{MN}(z)+h_{MN}(x,z)$,
with line element $ds^2=g_{MN}(z)dx^Mdx^N$ in a 5-dimensional anti-deSitter space or AdS$_5$,  in the double limit of large $N_c$ and strong gauge coupling~\cite{Nastase:2007kj}
(and references therein).
The form factors in (\ref{A1}) follow from the coupling  of the irreducible representations of the metric fluctuations  $h_{\mu\nu}$,  to a bulk Dirac fermion 
with chiral components $\psi_{L,R}$. The bulk metric fluctuations can be decomposed as~\cite{Kanitscheider:2008kd}.

\be
h_{\mu\nu}(k,z)\sim \epsilon_{\mu\nu}^{TT}h(k,z)+\frac{1}{3}\eta_{\mu\nu}f(k,z)
\ee
following  the $2\oplus 1\oplus 0$ invariant decomposition with  the spin-1 part omitted, as it drops from the threshold production analysis by parity.

\subsection{A-term}

To determine the A-term, we 
contract   the  energy-momentum form factor (\ref{A1}) with a spin-2 transverse-traceless polarization tensor $\epsilon_{\mu\nu}^{TT}$,

\bea
\label{EMT2}
\left<p_2|\epsilon_{\mu\nu}^{TT}T^{\mu\nu}(0)|p_1\right> &=&\overline{u}(p_2)\left(
A(k, \kappa_T)\epsilon_{\mu\nu}^{TT}\gamma^{\mu}p^{\nu}\right)u(p_1)\nonumber\\
\eea
and use a Witten diagram and  the holographic dictionary in the soft wall construction to evaluate it~\cite{Nastase:2007kj,Abidin:2009hr,Mamo:2019mka}.  
 (\ref{EMT2}) sources in bulk the  transverse and traceless part of the metric or a genuine spin-2 graviton (dual to $2^{++}$ tensor glueballs) 
 coupled to a bulk Dirac fermion, with the result~\cite{Abidin:2009hr,Mamo:2019mka}

\begin{widetext}
\bea
\label{AK11}
A(k,\kappa_T)=A(0)\times 6\times\frac{\Gamma \left(2+\frac{a_K}{2}\right)}{\Gamma \left(4+\frac{a_K}{2}\right)}\times \, _2F_1\left(3,\frac{a_K}{2};\frac{a_K}{2}+4;-1\right)\,,
\eea
\end{widetext}
with $a_K=K^2/4\kappa_T^2$ and $k^2=-K^2$ space-like. The scale $\kappa_T$ determines  the   dilaton profile in bulk,
and its value fixes most hadron and glueball trajectories  (see below). 
The hypergeometric function $_2F_1$ is related to the LerchPhi function.
In the large $N_c$ limit, $A(0)=1+\mathcal{O}(1/N_c)$ as the nucleon mass is totally glue dominated at low resolution. 
At finite $N_c$, only a fraction is glue dominated at the same resolution, so  $A(0)$ is a free parameter that can be fixed
by comparison to lattice results or experimental data.

\begin{figure*}
\subfloat[\label{fig_Alatticefit}]{%
  \includegraphics[height=5.5cm,width=.46\linewidth]{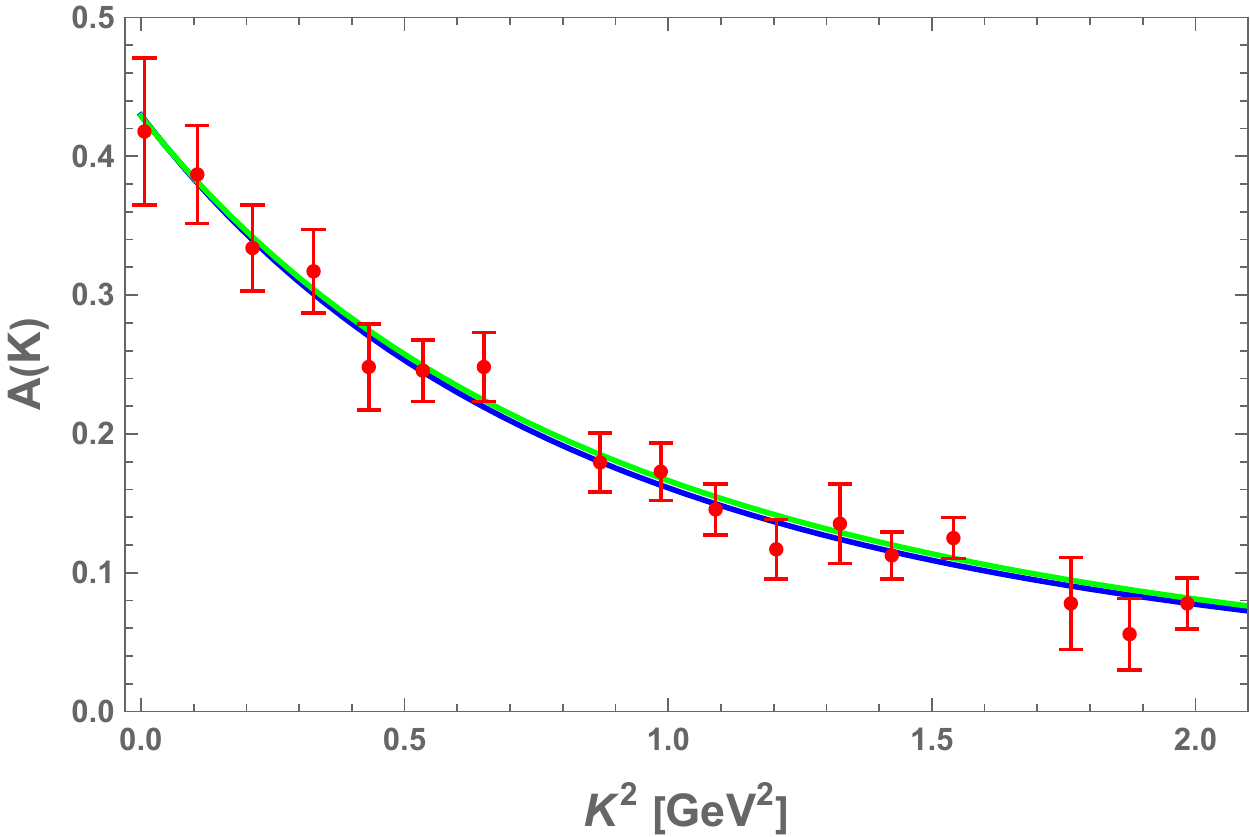}%
}\hfill
\subfloat[\label{fig_Dlatticefit}]{%
  \includegraphics[height=5.5cm,width=.46\linewidth]{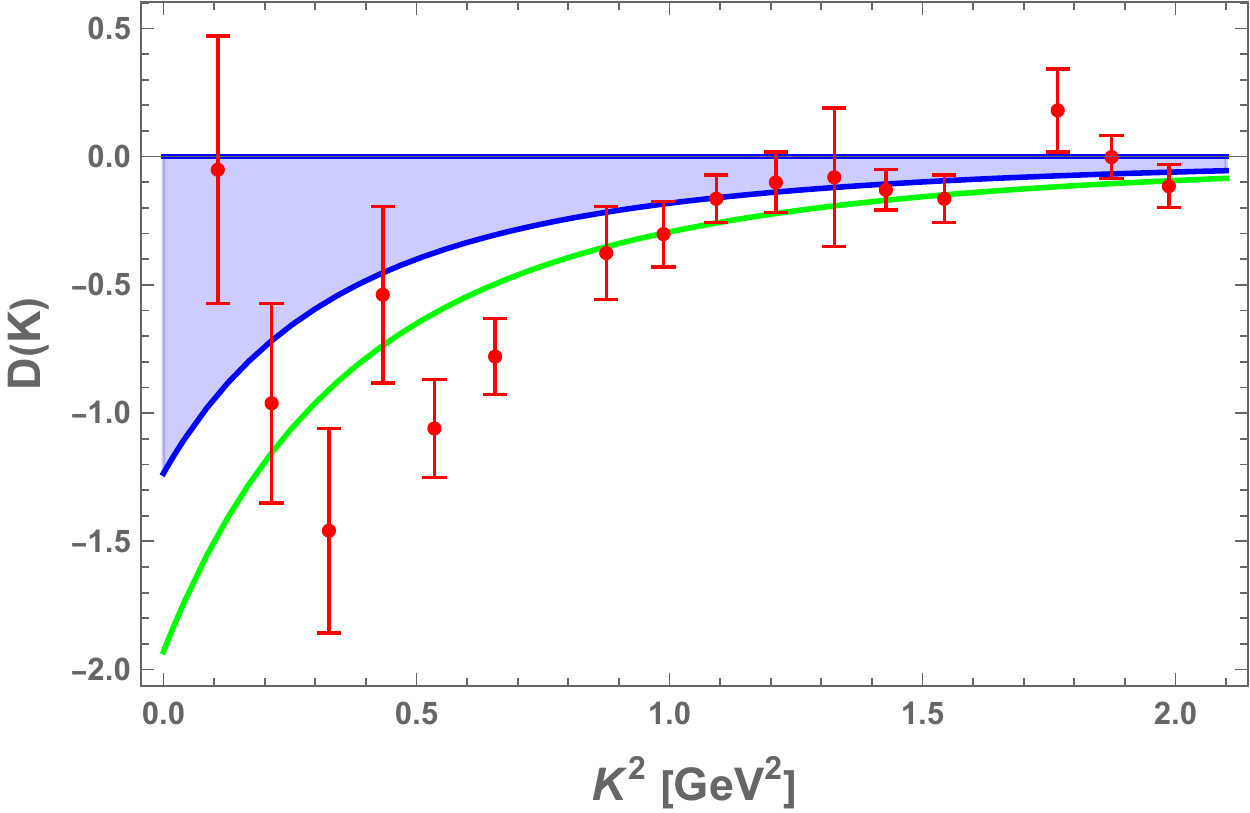}%
}
\caption{(a) The $A$ form factor from the recent lattice QCD result~\cite{Pefkou:2021fni} (red lattice data points), and our holographic fit using (\ref{AK11}) with $\kappa_T=0.388~\text{GeV}$, and $A(0)=0.430$ (solid-blue curve). The solid-green curve is the tripole lattice fit (\ref{Atripole}). (b) The $D$ form factor from the recent lattice QCD result~\cite{Pefkou:2021fni} (red lattice data points), and our holographic fit using (\ref{AS11}) with $\kappa_T=0.388~\text{GeV}$, $\kappa_S=0.217~\text{GeV}$, and $A(0)=0.430$ (lower blue-solid-curve). The blue shaded curve is determined by varying $\kappa_S$ between $\kappa_S=0.217~\text{GeV}$ (lower blue-solid-curve) with $D(0)=-1.275$,
and $\kappa_S=\kappa_T=0.388~\text{GeV}$ (upper blue-solid-curve) with $D(0)=0$ (large $N_c$ limit).
The solid-green curve is the tripole lattice fit (\ref{Atripole}).  See also Table \ref{table11}.}
  \label{fig_A-D_latticefit}
\end{figure*}

\subsection{C- or D-term}

To determine the C-term or  D-term ($D=4C$), we 
contract   the  energy-momentum form factor (\ref{A1}) with  $\frac{1}{3}\eta_{\mu\nu}$, 

\bea
\label{EMT22}
\frac{1}{3}\left<p_2|\eta_{\mu\nu}T^{\mu\nu}(0)|p_1\right> 
=\overline{u}(p_2)\left( A_S (k, \kappa_S)\frac{m_N}{3} \right)  u(p_1)\nonumber\\
\eea
and use again a Witten diagram for its holographic evaluation.
(\ref{EMT22}) sources in bulk  the trace part  of the metric (dual to $0^{++}$ tensor glueballs) coupled to a bulk Dirac fermion, with the result~\cite{Mamo:2021krl}
\bea
A_S(k, \kappa_S)=
A(k, \kappa_T)+\frac {k^2}{4m^2_N}B(k)-\frac{3k^2}{m^2_N}C(k, \kappa_T, \kappa_S)\nonumber\\
\eea
and 
\begin{widetext}
\bea
\label{AS11}
A_{S}(K,\kappa_S)&=&A(0)\times 6\times\frac{\Gamma \left(2+\frac{\tilde{a}_K}{2}\right)}{\Gamma \left(4+\frac{\tilde{a}_K}{2}\right)}\times \, _2F_1\left(3,\frac{\tilde{a}_K}{2};\frac{\tilde{a}_K}{2}+4;-1\right)\,,
\eea
with $\tilde{a}_K=K^2/4\kappa_S^2$. 
The B-term is kinematically excluded in holography, giving for $D=4C$
\bea
\label{A222}
D(k, \kappa_T, \kappa_S)=\frac{4}{3}\frac{m_N^2}{k^2}\bigg(A(k, \kappa_T)-A_S(k, \kappa_S)\bigg)
\eea
with $A_S(0)/A(0)=1$ by Poincare symmetry~\cite{Mamo:2021krl}. 
By expanding (\ref{A222}) near $k^2=0$, we can determine $D(0)$ as
\bea\label{DK0}
D(0)&\equiv&D(0,\kappa_T,\kappa_S) = -0.231\times A(0)\times m_N^2\times\left(\frac{1}{\kappa_S^2}-\frac{1}{\kappa_T^2}\right) \,.
\eea
\end{widetext}
(\ref{A222}) is the difference between the $2^{++}$ glueball Regge
trajectory and the $0^{++}$ glueball Regge trajectory, both of which are degenerate in leading order in $1/N_c$ (same anomalous
dimension) with $D(k, \kappa_T, \kappa_S)={\cal O}(1/N_c)$. The degeneracy is lifted at finite $1/N_c$ through different assignments for the anomalous dimensions~\cite{Gubser:2008yx}.
Since the leading scalar glueballs are lighter than the tensor glueballs, $D(k, \kappa_T, \kappa_S)$ is in general negative,  but small in magnitude~\cite{Mamo:2021krl}! 

\begin{widetext}
\begin{table}[h!]
\setlength{\arrayrulewidth}{0.1mm}
\setlength{\tabcolsep}{16pt}
\renewcommand{\arraystretch}{1.5}
\centering
\begin{tabular}{
 || m{2cm} 
 || m{0.4cm}
  | m{0.9cm}
  | m{0.9cm}
  | m{1.1cm}
  | m{1.1cm}
  | m{0.3cm}
  | m{0.6cm} 
  | m{1.7cm} || }
\hline\hline
                & $\tau$ & $\kappa_T(\text{GeV})$ & $\kappa_S(\text{GeV})$ & $m_{TT}(\text{GeV})$ & $m_{SS}(\text{GeV})$  & $A(0)$ & $D(0)$ & $\mathcal{N}e\,(\text{nb}\,{\rm GeV^{-2}})$ \\
\hline\hline
\textcolor{blue}{Holographic QCD $(This~work)$} & 3 & 0.388 & 0.217 (or 0.388) & \textbf{NA} & \textbf{NA} &  0.430 & $-1.275$ (or 0) & 2.032 (or 2.311 with $D(0)=0$) \\
\hline
\textcolor{blue}{Holographic QCD $(tripole~approx.)$} & \textbf{NA} & \textbf{NA} & \textbf{NA} &1.612 & 0.963 &  0.430 & $-1.275$ (or 0) & 2.032 (or 2.311 with $D(0)=0$) \\
\hline
\textcolor{Green}{Lattice QCD $(tripole~fit)$} ~\cite{Pefkou:2021fni} & \textbf{NA} &  \textbf{NA} & \textbf{NA} & 1.641 & 1.070 &  0.429 & $-1.930$  & 2.549
\\
\hline
\textcolor{Purple}{Lattice QCD $(dipole~fit)$}~\cite{Shanahan:2018pib} & \textbf{NA} &  \textbf{NA} & \textbf{NA} & 1.130 & 0.480 &  0.580 & $-10.000$  & 3.244  
\\ 
\hline\hline
\end{tabular}
\caption{Summary of the values we have used for the holographic QCD parameters of the gravitational form factors, and the corresponding normalization constant of the differential cross section (\ref{diff1}) (which are $\tau$, $\kappa_T$, $\kappa_S$, $A(0)$, 
$\mathcal{N}e$, as D(0) is totally fixed), and the lattice QCD tripole (\ref{Atripole})~\cite{Pefkou:2021fni}  and dipole fit (\ref{Atripole})~\cite{Shanahan:2018pib} parameters of the gravitational form factors, and the corresponding normalization constant of the differential cross section (\ref{diff1}) (which are $m_{TT},m_{SS}$, $A(0)$, $D(0)$, and $\mathcal{N}e$). Note that $\textbf{NA}$ is a short hand for not available (i.e., the parameter is not part of the corresponding model fit).}\label{table11}
\end{table}

\subsection{Comparison to the new lattice results}

The parameter $\kappa_T$ is related to the dilaton profile in the soft wall model, and its 
value is fixed by the rho meson Regge trajectory $\kappa_T=m_\rho/2=0.388~\text{GeV}$. Note that this
value fixes rather well most of  the electromagnetic radii of the nucleon~\cite{Mamo:2021jhj}. The value
of the tensor gravitational form factor $A(0)=1+{\cal O}(1/N_c)$  is only fixed at asymptotic $N_c$. At
finite $N_c$, a comparison with the recent lattice results~\cite{Pefkou:2021fni} suggests $A(0)=0.430$.
We fix the parameter  $\kappa_S=\kappa_T(1+{\cal O}(1/N_c))$ in two ways: 1/ At large $N_c$ with $\kappa_S=\kappa_T$,
for which the D form factor is null; 2/ At finite $N_c$ with $\kappa_S=0.217$ GeV,  so that the
holographic $D$ form factor (\ref{DK2}) is comparable to the recent lattice  $D$ form factor~\cite{Pefkou:2021fni}. For numerical convenience, we can also approximate the holographic A and D form factors (\ref{AK11} and \ref{A222}) by a tripole form as
\bea\label{ADtripoleApprox}
A(K,m_{TT})&=&\frac{A(0)}{\left(1+\frac{K^2}{m_{TT}^2}\right)^3}\,,\nonumber\\
D(K,m_{SS})&=&\frac{D(0)}{\left(1+\frac{K^2}{m_{SS}^2}\right)^3}\,,
\eea
with $m_{TT}=1.612~\text{GeV}$, $A(0)=0.430$ and 
 $m_{SS}=0.963~\text{GeV}$,  $D(0)=-1.275$, respectively, see also Table \ref{table11}.

For comparison, we quote the tripole fit to the lattice data for the A and D form factors given in~\cite{Pefkou:2021fni}

\bea\label{Atripole}
A(K,m_{TT})&=&\frac{A(0)}{\left(1+\frac{K^2}{m_{TT}^2}\right)^3}\,,\nonumber\\
D(K,m_{SS})&=&\frac{D(0)}{\left(1+\frac{K^2}{m_{SS}^2}\right)^3}\,,
\eea
with $m_{TT}=1.641~\text{GeV}$, $A(0)=0.429$ and 
 $m_{SS}=1.070~\text{GeV}$,  $D(0)=-1.930$ respectively.
Also and for completeness, 
the dipole  fit to the lattice data for the A and D form factors given in~\cite{Shanahan:2018pib}
\be\label{Adipole}
A(K,m_{TT})&=&\frac{A(0)}{\left(1+\frac{K^2}{m_{TT}^2}\right)^2}\,,\nonumber\\
D(K,m_{SS})&=&\frac{D(0)}{\left(1+\frac{K^2}{m_{SS}^2}\right)^2}\,,
\eea
with $m_{TT}=1.130~\text{GeV}$, $A(0)=0.580$, and 
$m_{SS}=0.480~\text{GeV}$, and $D(0)=-10$ respectively.

In  Fig.~\ref{fig_Alatticefit}, we show the A gravitational form factor versus $K^2$.
The solid-blue line is our holographic prediction with the newly adjusted value $A(0)=0.430$,
the green-solid line is the tripole lattice fit (\ref{Atripole}). The red lattice data points are from~\cite{Pefkou:2021fni}.
The holographic results is a total match. 
In  Fig.~\ref{fig_Dlatticefit}, we show the D gravitational form factor versus $K^2$.
The holographic result is indicated by the blue-band which is obtained by varying
$\kappa_S$ between $\kappa_S=0.217~\text{GeV}$ (lower blue-solid-curve) with $D(0)=-1.275$,
and $\kappa_S=\kappa_T=0.388~\text{GeV}$ (upper blue-solid-curve) with $D(0)=0$ (large $N_c$ limit).
The solid-green curve is the tripole lattice fit (\ref{Atripole}), and the lattice data are from~\cite{Pefkou:2021fni}.
A comparison of the parameters entering the holographic approach used in this work, and the ones we used in an earlier
work~\cite{Mamo:2021krl} are recorded in Table \ref{table11}.  We have also noted the changes in the lattice parameters reported in~\cite{Pefkou:2021fni}
from those reported earlier in~\cite{Shanahan:2018pib}.

\begin{table}[h!]
\setlength{\arrayrulewidth}{0.1mm}
\setlength{\tabcolsep}{16pt}
\renewcommand{\arraystretch}{1.5}
\centering
\begin{tabular}{
 || m{2cm} 
  | m{4cm}
  | m{4cm} || }
\hline\hline
                & $\langle r^2_{GS}\rangle~(\text{fm}^2)$ & $\langle r^2_{GM}\rangle~(\text{fm}^2)$  \\
\hline\hline
\textcolor{blue}{Holographic QCD $(This~work)$} &  $\frac{1.04}{\kappa_S^2}$\,=\,$0.926^2$  ($or~0.518^2~with$ $\kappa_S=\kappa_T$ ) & $\frac{0.693}{\kappa_T^2}$\,+\,$\frac{0.347}{\kappa_S^2}$\,=\,$0.682^2$ ($or~0.518^2~with$ $\kappa_S=\kappa_T$ ) \\
\hline
\textcolor{blue}{Holographic QCD $(tripole~ approx.)$} & $-\frac{D(0)}{A(0)}\times\frac{4.5}{m_N^2}+\frac{18}{m_{TT}^2}=0.926^2$ &$-\frac{D(0)}{A(0)}\times\frac{1.5}{m_N^2}+\frac{18}{m_{TT}^2}=0.682^2$ \\
\hline
\textcolor{Green}{Lattice QCD $(tripole fit)$} ~\cite{Pefkou:2021fni} &  $-\frac{D(0)}{A(0)}\times\frac{4.5}{m_N^2}+\frac{18}{m_{TT}^2}= 1.073^2$ & $-\frac{D(0)}{A(0)}\times\frac{1.5}{m_N^2}+\frac{18}{m_{TT}^2}=0.747^2$ \\
\hline
\textcolor{Purple}{Lattice QCD $(dipole fit)$}~\cite{Shanahan:2018pib} & $-\frac{D(0)}{A(0)}\times\frac{4.5}{m_N^2}+\frac{12}{m_{TT}^2}=1.945^2$ & $-\frac{D(0)}{A(0)}\times\frac{1.5}{m_N^2}+\frac{12}{m_{TT}^2}=1.227^2$  
\\ 
\hline\hline
\end{tabular}
\caption{Summary of the gluonic scalar (GS), and gluonic mass (GM) radii (\ref{ASR} and \ref{AMR}) with the holographic gravitational form factors (\ref{AK11} and \ref{A222}) (including their tripole approximation (\ref{ADtripoleApprox})), and the lattice QCD tripole (\ref{Atripole})~\cite{Pefkou:2021fni} and dipole fits (\ref{Adipole})~\cite{Shanahan:2018pib} of the lattice gravitational form factor data.}\label{table112}
\end{table}

\begin{figure*}
\subfloat[\label{fig_RGM}]{%
  \includegraphics[height=5.5cm,width=.46\linewidth]{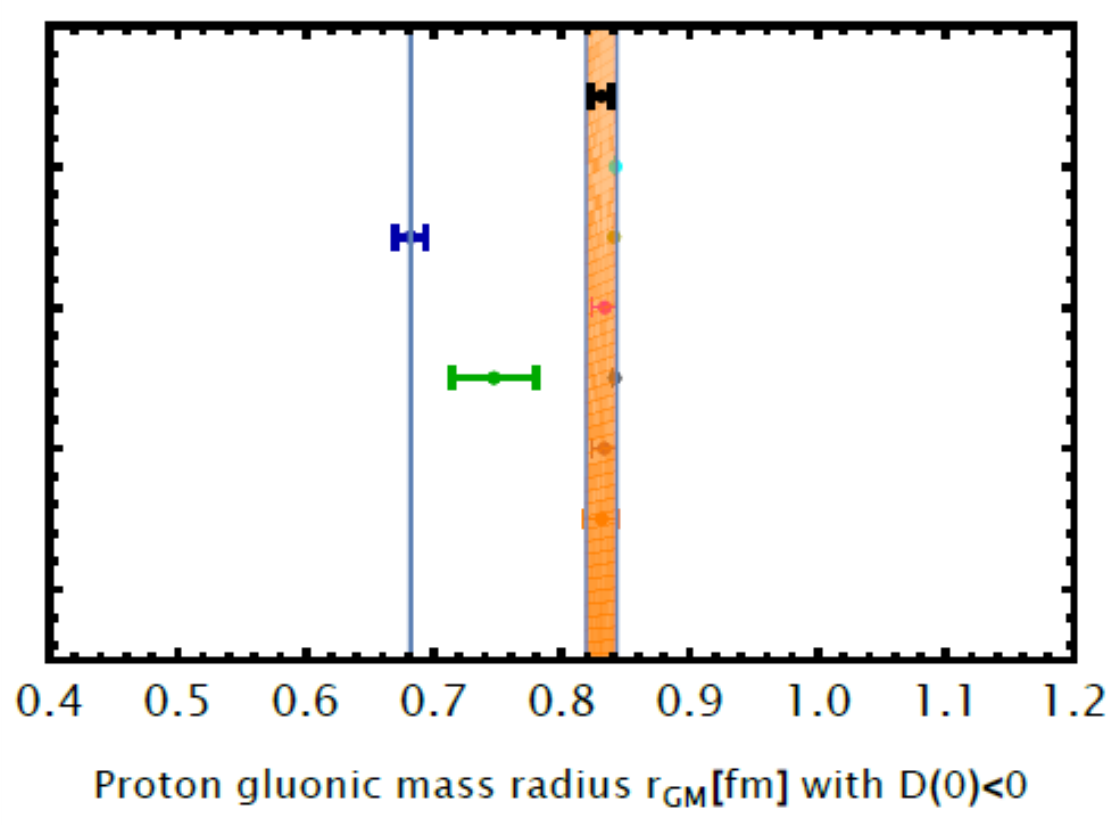}%
}\hfill
\subfloat[\label{fig_RGS}]{%
  \includegraphics[height=5.5cm,width=.46\linewidth]{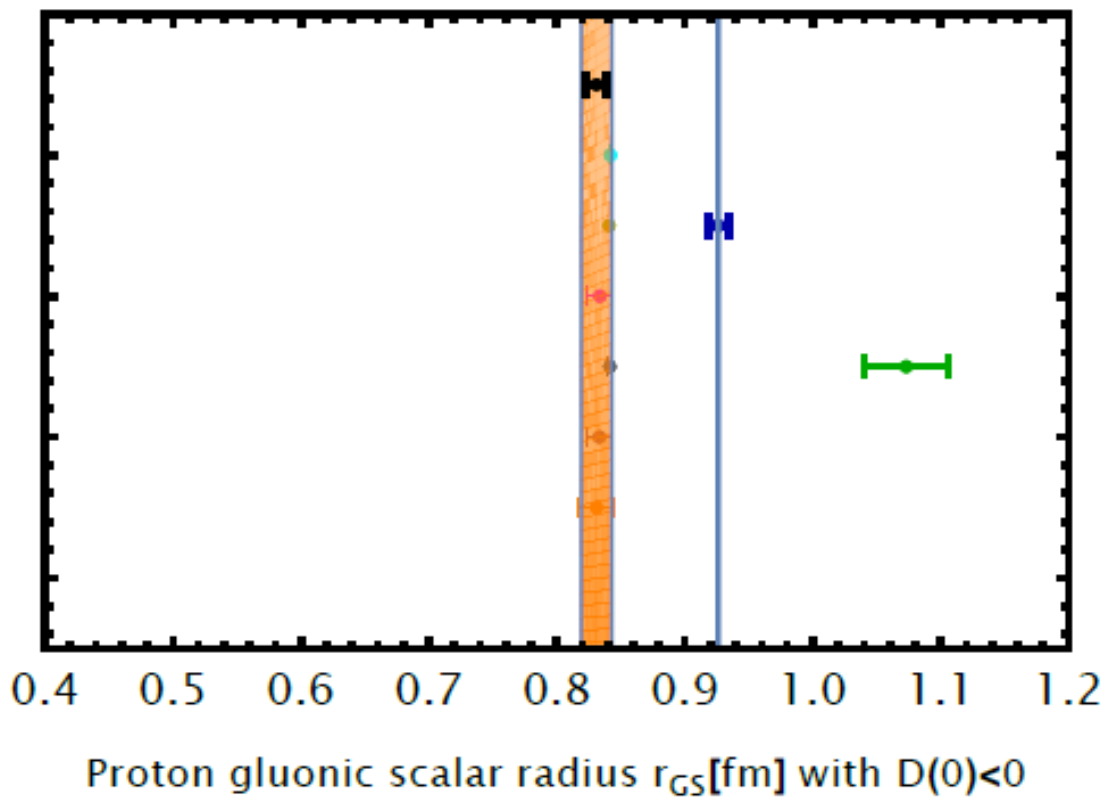}%
}
\caption{(a) The gluonic mass radius $r_{GM}$ defined in (\ref{AMR}). The blue data point at $r_{GM}=0.682\pm 0.012~\rm fm$ is determined by using our holographic A and D form factors (\ref{AK11} and \ref{A222}), and the green data point at $r_{GM}=0.747\pm 0.033~\rm fm$ is determined using the lattice tripole fit to the lattice A and D form factor data (\ref{Atripole})~\cite{Pefkou:2021fni}.  (b) The gluonic scalar radius $r_{GS}$ defined in (\ref{ASR}). The blue data point at $r_{GS}=0.926\pm 0.008~\rm fm$ is determined by using our holographic A and D form factors (\ref{AK11} and \ref{A222}), and the green data point at $r_{GS}=1.073\pm 0.033~\rm fm$ is determined by using the lattice tripole fit to the lattice A and D form factor data (\ref{Atripole})~\cite{Pefkou:2021fni}. For comparison, we have also shown the charge radius of the proton (the orange band) centered around the recent PRad measurement of $r_{c}=0.831~\rm fm$ \cite{Xiong:2019umf}. The black data point is our holographic prediction of the charge radius of proton $r_{c}=0.831\pm 0.008~\rm fm$ \cite{Mamo:2021jhj}. The other charge radius data points, within the orange band, are from Pohl 2010 ($\mu$ H spect.) \cite{Pohl:2010zza}, Antognini 2013 ($\mu$ H spect.) \cite{Antognini:2013txn}, Beyer 2017 (H spect.) \cite{CREMA_2017}, CODATA 2018 \cite{CODATA_2018}, and Bezignov 2019 (H spect.) \cite{Bezginov:2019mdi}.}
  \label{fig_A-D_latticefit}
\end{figure*}


\subsection{Radii, pressure, and shear distributions}

\subsubsection{Radii}
Given the gravitational A and D form factors, a number of radii can be defined. The simplest are those directly given by A and D respectively,
which characterize the tensor and scalar glueball range as discussed in~\cite{Mamo:2021krl}. Alternatively, we may follow~\cite{Ji:2021mtz} and define 
the gluonic scalar mass radius $r_{GS}$ of the proton (derived from the trace of its energy-momentum tensor (\ref{A1}) with $k^2=-K^2$)  as 
\be
\label{ASR}
\langle r^2_{GS}\rangle =-\frac{6}{A_S(0)}\bigg(\frac{d A_S(K)}{dK^2}\bigg)_0\,\hbar^2 c^2\,,
\ee
where
\bea
A_S(K)\equiv
A(K)-\frac {K^2}{4m^2_N}B(K)+\frac{3K^2}{4m^2_N}D(K)\,.\nonumber\\
\eea
Similarly, the gluonic mass radius $r_{GM}$ of the proton (derived from the $00$ component of its energy-momentum tensor (\ref{A1}) with $k^2=-K^2$) 
is defined as~\cite{Ji:2021mtz}
\be
\label{AMR}
\langle r^2_{GM}\rangle =-\frac{6}{A_M(0)}\bigg(\frac{d A_M(K)}{dK^2}\bigg)_0\,\hbar^2 c^2\,,
\ee
where
\be
A_{M}(K)\equiv A(K)-\frac {K^2}{4m^2_N}B(K)+\frac{K^2}{4m^2_N}D(K)\,.\nonumber\\
\ee
Throughout, we use $\hbar c=0.197\,\text{GeV}\,\rm fm$. See Table \ref{table112}, for comparison of the gluonic radii of the proton that we have found by using the holographic (\ref{AK11} and \ref{A222}) and lattice tripole or dipole fits (\ref{Atripole} or \ref{Adipole}) of the $A$ and $D$ gravitational form factors in (\ref{ASR}) and (\ref{AMR}). Note that all the radii in Table \ref{table112} are computed with $B(K)=0$.

In Fig.~\ref{fig_RGM}, we show the gluonic mass radius $r_{GM}$ defined in (\ref{AMR}). The blue data point at $r_{GM}=0.682\pm 0.012~\rm fm$ is determined by using our holographic A and D form factors (\ref{AK11} and \ref{A222}), and the green data point at $r_{GM}=0.747\pm 0.033~\rm fm$ is determined using the lattice tripole fit to the lattice A and D form factor data (\ref{Atripole})~\cite{Pefkou:2021fni}.  In Fig.~\ref{fig_RGS}, we show  the gluonic scalar radius $r_{GS}$ defined in (\ref{ASR}). The blue data point at $r_{GS}=0.926\pm 0.008~\rm fm$ is determined by using our holographic A and D form factors (\ref{AK11} and \ref{A222}), and the green data point at $r_{GS}=1.073\pm 0.033~\rm fm$ is determined by using the lattice tripole fit to the lattice A and D form factor data (\ref{Atripole})~\cite{Pefkou:2021fni}. Note that the gluonic scalar
radius is smaller for $D(0)=0$ (strict large $N_c$ limit  with $\kappa_S=\kappa_T$) with $r_{GS}=r_{GM}=0.518\pm 0.018~\rm fm$, determined  using our holographic A form factor (\ref{AK11}).
The theoretical uncertainty for the holographic results are determined by varying $\kappa_T=0.388~\text{GeV}$ to $\kappa_T=0.402~\text{GeV}$ (which gives a charge radius of the proton to be $0.831~\rm fm$ \cite{Mamo:2021jhj} (in perfect agreement with the recent PRad measurement \cite{Xiong:2019umf})). We have also estimated the error in the lattice results to be $\pm 0.033$, see Table X in \cite{Pefkou:2021fni}.  See also Table \ref{table112}.

\subsubsection{Pressure and shear distributions}

In QCD, the proton, as an extended quantum object, is composed of  interacting partons in different proportions of
quarks and gluons,  at different resolutions. This composition at the nucleon mass resolution,  is clearly non-perturbative.
At large $N_c$ and fixed $^\prime$t Hooft coupling $\lambda =g^2N_c$, this nonperturbative description is captured by 
 interacting and topological pion and vector meson fields, in the form of a  Skyrmion~\cite{Zahed:1986qz}. The meson mediated force field is repulsive at the core,
and attractive at the periphery, setting the pressure force across the nucleon~\cite{Jung:2013bya}.

Holography is yet another dual and nonperturbative description at large $N_c$ and large $^\prime$t Hooft coupling $\lambda =g^2N_c$, 
whereby the nucleon hologram is held together by gravitons (dual of tensor glueballs)
and possibly scalars (dual of scalar glueballs) emerging from bulk string fluctuations. The short range tensor gravitons are dominant at the core and repulsive, while the 
long range scalars are attractive at the periphery~\cite{Mamo:2021krl}.
Using the analytic Fourier transform of a tripole D form factor (with $-t=K^2=\vec{\Delta}^2$)~\cite{Lorce:2018egm,Pefkou:2021fni}
\bea
\label{DR}
\tilde{D}(r)&=&\int \frac{d^3\vec{\Delta}}{(2\pi)^3}\,e^{-i\vec{\Delta}\cdot\vec{r}}\,D(\vec{\Delta},m_{SS})
=\int \frac{d^3\vec{\Delta}}{(2\pi)^3}\,e^{-i\vec{\Delta}\cdot\vec{r}}\,\frac{D(0)}{\left(1+\frac{\vec{\Delta}^2}{m_{SS}^2}\right)^3}=D(0)\times\frac{m_{SS}^3}{32\pi}\times (1+m_{SS}r) \times e^{-m_{SS}r}\,,\nonumber\\
\eea
of the holographic tripole approximation (\ref{ADtripoleApprox}), and lattice tripole fits (\ref{Atripole}), we can determine the gluonic pressure and shear distributions inside the proton (in the Breit frame with $\vec{p}=0$)
\cite{Polyakov:2002yz,Polyakov:2018zvc} 
\bea
\label{p}
p(r)=&&\frac{1}{6m_N} \frac{1}{r^2}\frac{d}{dr} \bigg(r^2\frac{d}{dr}
{\tilde{D}(r)}\bigg)\,,\\
s(r)= &&-\frac{1}{4m_N} r\frac{d}{dr}\bigg( \frac{1}{r} \frac{d}{dr}
{\tilde{D}(r)}\bigg)\,.\label{s}
\eea
The pressure and shear distributions capture the anisotropic spatial content of the energy momentum tensor as
\be
\label{TIJ}
T^{ij}(\vec r)=\frac 13 \delta^{ij} p(r)+\bigg(\hat{r}^i\hat{r}^j-\frac 13 \delta^{ij}\bigg)s(r)\,.
\ee

In Fig.~\ref{fig_PR}, we show the pressure distribution $\hbar^2c^2\times r^2p(r)$ inside proton (\ref{p} with \ref{DR}) where $m_{ss}=0.963~\text{GeV}/\hbar c$ and $D(0)=-1.275$ (solid-blue curve (holographic QCD)), and $m_{ss}=1.070~\text{GeV}/\hbar c$ and $D(0)=-1.930$ (solid-green curve (lattice QCD)). 
In Fig.~\ref{fig_SR}, we show the shear distribution  $\hbar^2c^2\times r^2s(r)$ inside proton (\ref{s} with \ref{DR}) where $m_{ss}=0.963~\text{GeV}/\hbar c$ (solid-blue curve (holographic QCD)), and $m_{ss}=1.070~\text{GeV}/\hbar c$ (solid-green curve (lattice QCD)). The shaded blue curves in (a) and (b) correspond to varying the holographic $D(0)$ value of the form factor between $-1.275$ and $0$. For the latter, the scalar and tensor glueball Regge trajectories become degenerate, balancing identically each other,
with zero pressure across the proton!

\begin{figure*}
\subfloat[\label{fig_PR}]{%
  \includegraphics[height=5.5cm,width=.46\linewidth]{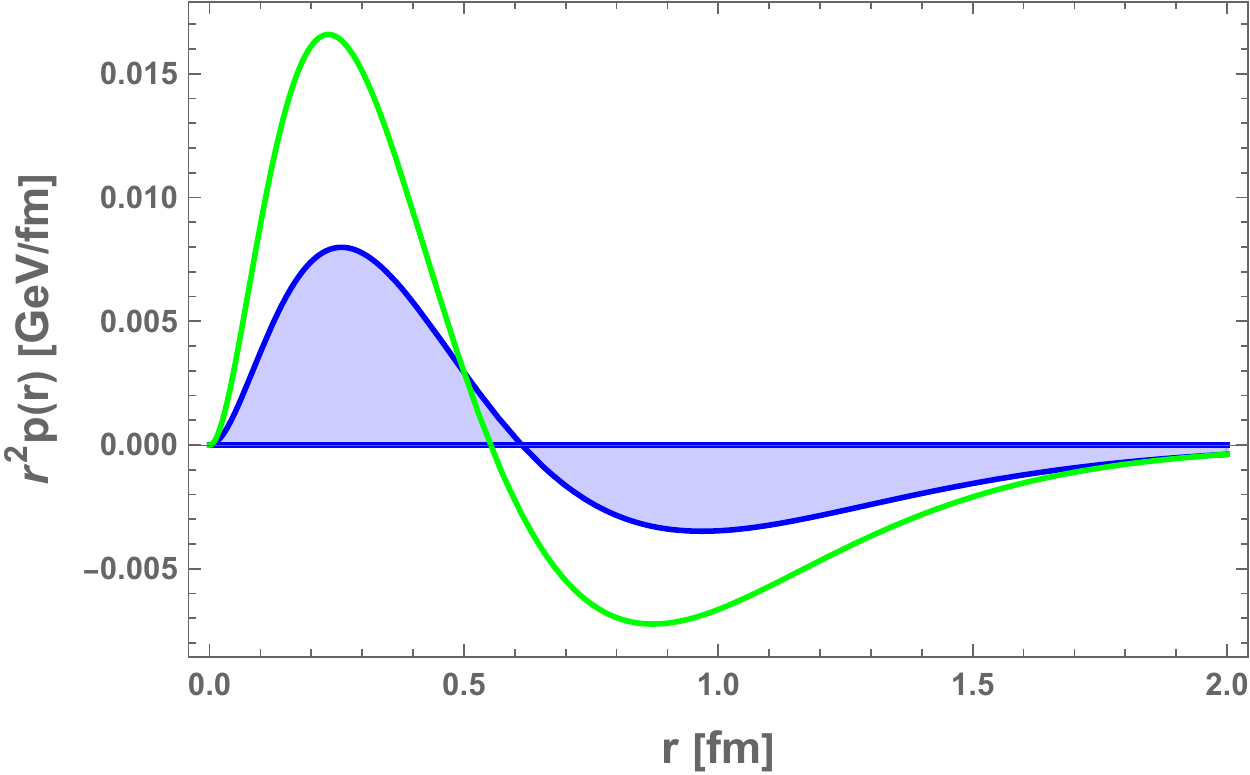}%
}\hfill
\subfloat[\label{fig_SR}]{%
  \includegraphics[height=5.5cm,width=.46\linewidth]{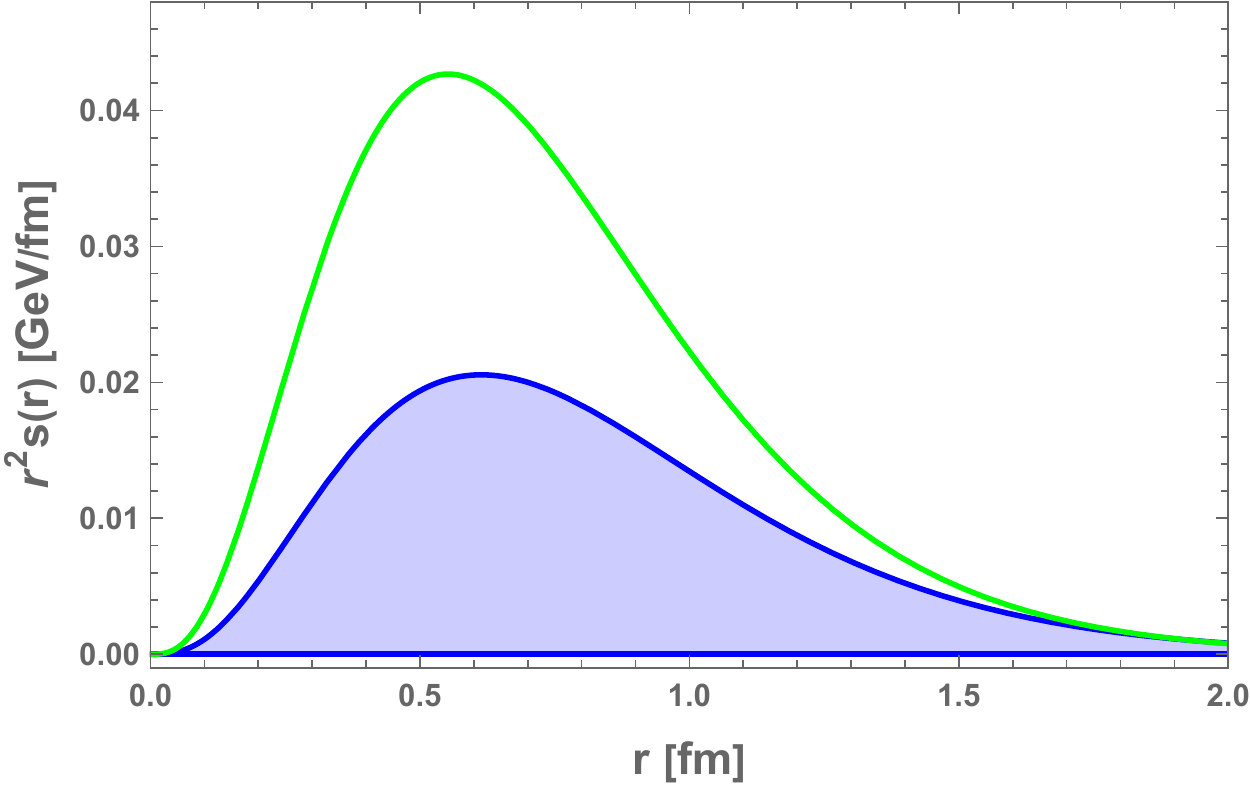}%
}
\caption{(a) The pressure distribution $\hbar^2c^2\times r^2p(r)$ inside the proton (\ref{p} with \ref{DR}) where $m_{ss}=0.963~\text{GeV}/\hbar c$ and $D(0)=-1.275$ (solid-blue curve (holographic QCD)), and $m_{ss}=1.070~\text{GeV}/\hbar c$ and $D(0)=-1.930$ (solid-green curve (lattice QCD)). (b) The shear distribution  $\hbar^2c^2\times r^2s(r)$ inside proton (\ref{s} with \ref{DR}) where $m_{ss}=0.963~\text{GeV}/\hbar c$ (solid-blue curve (holographic QCD)), and $m_{ss}=1.070~\text{GeV}/\hbar c$ (solid-green curve (lattice QCD)). The shaded blue curves in (a) and (b) correspond to varying the holographic $D(0)$ between $-1.275$ and $0$.}
  \label{fig_pressure-shear}
\end{figure*}

\section{Holographic  $J/\psi$ production near-threshold}~\label{EE}

The A and to a lesser extent 
D gravitational  form factors, are the key elements in  the electromagnetic production of heavy mesons in the threshold region.  The holographic approach allows
not only for their determination as we briefly reviewed, but fixes the entire  heavy vector meson production process time-like from threshold to asymptotia,  all within one consolidated framework. We now briefly detail the electroproduction process for charmonia near threshold and close to the photon point, and use that to reanalyze the GlueX data, for a better
understanding of the role of the D term.

\subsection{Holographic electroproduction}

The t-channel $2^{++}$ glueball exchange contribution to the electroproduction amplitude of vector mesons  is described in bulk by the exchange of a graviton~\cite{Mamo:2021tzd}.
The Witten amplitude for the transverse and longitudinal ($T,L$) electroproduction of a heavy meson ($V$) off a nucleon, are

\bea
\label{Amph0}
&&i{\cal A}^{h}_{\gamma^{*}_{T/L}p\rightarrow  Vp} (s,t)=\sum_n i\tilde{{\cal A}}^{h}_{\gamma^{*}_{T/L}p\rightarrow  Vp} (m_n,s,t)\nonumber\\
&&i\tilde{{\cal A}}^{h}_{\gamma_{T/L}^{*}p\rightarrow  Vp} (m_n,s,t)=\frac{1}{g_5}\times(-i)V_{h\gamma^{*}_{T/L}V}^{\mu\nu}(q,q^{\prime},k,m_n)\times \tilde{G}_{\mu\nu\alpha\beta}(m_n,k)\times
 (-i)V_{h\bar\Psi\Psi}^{\alpha\beta}(p_1,p_2,k,m_n)\,,	\nonumber\\
\eea
with the bulk vertices (defining $t=k^2=\Delta^2$, and $k=\Delta=p_2-p_1=q-q^{\prime}$)

\be
\label{Amph1}
&&V_{h\gamma^{*}_{T/L}V}^{\mu\nu}(q,q^{\prime},k,m_n)\equiv \left(\frac{\delta S_{h\gamma^{*}_{T/L}V}^k}{\delta (\epsilon_{\mu\nu}h(k,z))}\right)\,J_{h}(m_n,z)=\sqrt{2\kappa^2}\times\frac{1}{2}\int dz\sqrt{g}\,e^{-\phi}z^4K_{T/L}^{\mu\nu}(q,q^{\prime},\epsilon,\epsilon^{\prime},z)J_{h}(m_n,z)\,,\nonumber\\
&&V_{h\bar\Psi\Psi}^{\alpha\beta}(p_1,p_2,k,m_n)\equiv \left(\frac{\delta S_{h\bar\Psi\Psi}^k}{\delta (\epsilon_{\alpha\beta}h(k,z))}\right)\,J_{h}(m_n,z)=-\sqrt{2\kappa^2}\times\frac{1}{2}\int dz\sqrt{g}\,e^{-\phi}z\bar\Psi(p_2,z)\gamma^\alpha p^\beta\Psi(p_1,z)J_{h}(m_n,z)\,.\nonumber\\ \label{vh}
\ee
The graviton bulk-to-bulk propagator $\tilde G$, and the U(1) vector meson bulk-to-boundary transverse and longitudinal
propagators $K_{T/L}$ are given  in Appendix~\ref{APP_EE}.  The holographic wavefunction for the spin-2 glueball $J_h(m_n,z)$ is given
in Appendix~\ref{APP_GB} for the soft wall model, and the holographic wavefunction $\Psi(p,z)$ refers to the bulk Dirac fermion. The details
regarding the reduction of (\ref{Amph0}-\ref{Amph1}) can be found in Appendix~\ref{APP_EE}, with the reduced transverse and longitudinal amplitudes

\bea
\label{nAmphfull2}
{\cal A}^{h}_{\gamma_{T}^{*}p\rightarrow Vp} (s,t) &=& -\frac{1}{g_5}\times 2\kappa^2\times\mathcal{V}_{h\gamma_T^{*}V}(Q,M_V)\times\frac{1}{m_N}\times\bar u(p_2)u(p_1)\nonumber\\
&\times & \left(p_1\cdot q\right)^{2}\times\left[A_T(K,\kappa_T)+\eta^2 D(K,\kappa_T,\kappa_S)\right]\,,\nonumber\\
{\cal A}^{h}_{\gamma_{L}^{*}p\rightarrow Vp} (s,t) &=& -\frac{1}{g_5}\times 2\kappa^2\times\frac{1}{QM_V}\times\mathcal{V}_{h\gamma_L^{*}V}(Q,M_V)\times\frac{1}{m_N}\times\bar u(p_2)u(p_1)\nonumber\\
&\times & \left(p_1\cdot q\right)^{2}\times\left[A_T(K,\kappa_T)+\eta^2 D(K,\kappa_T,\kappa_S)\right]\,,
\eea
with the A and D gravitational form factors following from
\bea\label{AK1}
A(K,\kappa_T)&\equiv&\frac{1}{2}\int dz\,\sqrt{g}\,e^{-\phi}z\,\left(\psi_R^2(z)+\psi_L^2(z)\right)\times\sum_{n=0}^{\infty}\frac{\sqrt{2}\kappa F_{n}\psi_n(z)}{K^2+m_n^2}\,,\nonumber\\
&=&\frac{1}{2}\int dz\,\sqrt{g}\,e^{-\phi}z\,\left(\psi_R^2(z)+\psi_L^2(z)\right)\times\mathcal{H}(K,z)\,,\nonumber\\
&=&A(0)\times 6\times\frac{\Gamma \left(2+\frac{a_K}{2}\right)}{\Gamma \left(4+\frac{a_K}{2}\right)}\times \, _2F_1\left(3,\frac{a_K}{2};\frac{a_K}{2}+4;-1\right)\,,
\eea
and
\bea\label{DK1}
D(K,\kappa_T)&\equiv&\frac{1}{2}\int dz\,\sqrt{g}\,e^{-\phi}z\,\left(\psi_R^2(z)+\psi_L^2(z)\right)\times\sum_{n=0}^{\infty}\,\frac{4}{3}\frac{p_1^2}{m_n^2}\times\frac{\sqrt{2}\kappa F_{n}\psi_n(z)}{K^2+m_n^2}\,,\nonumber\\
&=&\frac{1}{2}\int dz\,\sqrt{g}\,e^{-\phi}z\,\left(\psi_R^2(z)+\psi_L^2(z)\right)\times\sum_{n=0}^{\infty}\,\frac{4}{3}\frac{p_1^2}{k^2}\times\frac{\sqrt{2}\kappa F_{n}\psi_n(z)}{-k^2+m_n^2}\,,\nonumber\\
&=&-\frac{4}{3}\frac{m_N^2}{K^2}\times A(K,\kappa_T)\,,
\eea
\end{widetext}
with the dilaton profile $\phi(z)=\kappa_T^2 z^2$. In deriving the last two relations in (\ref{DK1}), we have assumed that the D form factor is analytic in $K^2$, and traded $m_n^2\rightarrow k^2$ at the pole time-like, before
switching back space-like.  Note that  only the A form factor contribution appears in (\ref{DK1}), as expected from a graviton exchange.  The complementary trace part contribution due to the exchange of a dilaton in bulk, fails to couple to a heavy vector meson described by a U(1) bulk gauge field~\cite{Mamo:2021krl}. 
Since we cannot rule out a holographic formulation with a heavy vector meson in bulk with a tracefull coupling at finite $N_c$, we will assume~\cite{Mamo:2021krl}

\begin{widetext}
\bea\label{DK2}
D(K, \kappa_T)\rightarrow D(K,\kappa_T,\kappa_S)&=&-\frac{4}{3}\frac{m_N^2}{K^2}\times \left[A(K,\kappa_T)-A_{S}(K,\kappa_S)\right]\,.
\eea

\end{widetext}

\subsection{Holographic photoproduction}


The coupling between $2^{++}$ glueballs (denoted by $h$), vector mesons with mass $M_V$ (denoted by $V$), and transverse virtual photons with $Q^2\neq0$ (denoted by $\gamma_T^*$) or transverse real photons with $Q^2=0$ (denoted by $\gamma_T$)  is given by (\ref{couplingT})

\be \label{vvJPSI2soft}
\mathcal{V}_{h\gamma_T V} &\equiv&\mathcal{V}_{h\gamma^{*}_TV}(Q=0,M_V) = g_5\times \bigg(\frac {f_V}{M_V}\bigg)\times{\mathbb V}_{h\gamma V}\,,\nonumber\\
\ee
where $f_V/M_V$ is the dimensionless ratio of the decay constant $f_V$ of $V=J/\psi$, and its mass $M_V$. Here we have also defined
\be\label{GlueballMesonCoupling}
{\mathbb V}_{h\gamma V}=\frac{1}{2}\frac{1}{4\tilde{\kappa}_{V}^4}\int_{0}^{\infty} d\tilde{\xi} \,e^{-\tilde{\xi}}\,\tilde{\xi}^2\times L_{0}^1(\tilde{\xi}) =\frac{1}{4\tilde{\kappa}_{V}^4}\,.\nonumber\\
\ee
The pertinent differential cross section at the photon point,  is of the form

\be\label{DCC}
\left(\frac{d\sigma}{dt}\right)
=\frac{e^2}{16\pi(s-m_N^2)^2}\,\frac 12\sum_{{\rm T=1,2}}
\frac 12\sum_{{\rm spin}}
\Bigg|{\cal A}^{h}_{\gamma_T p\rightarrow  J/\Psi p} (s,t)\Bigg|^2\,,\nonumber\\
\ee
which is dominated by the TT-part of the graviton or $2^{++}$ glueball exchange as we noted earlier.
The first sum is over the photon and $J/\Psi$ polarizations $T=1,2$.
 The second sum is over the initial and final bulk Dirac fermion (proton) spin, evaluated explicitely as
\begin{widetext}
\bea
\sum_{s,s'}\bar u_{s'}(p_2)u_s(p_1)\bar u_s(p_1)u_{s'}(p_2)&=& \Tr\Big(\sum_{s.s'}u_{s'}(p_2)\bar u_{s'}(p_2)u_s(p_1)\bar u_s(p_1)\Big)\nonumber\\
&=&\frac 14 \Tr\Big(\big(\gamma_\mu p_2^\mu+m_N\big)\big(\gamma_\mu p_1^\mu+m_N\big)\Big)=8m_N^2\times\left(1+\frac{K^2}{4m_N^2}\right)\approx 8m_N^2\,,\nonumber\\
\eea
for $\frac{K^2}{4m_N^2}\ll1$ or $p_1\sim p_2$ which we are assuming in the high energy limit $s\gg -t$.

Carrying explicitly these summations yield the differential cross section for photoproduction of heavy meson in the spin $j=2$
exchange approximation as

\be
\label{diff1}
\left(\frac{d\sigma}{dt}\right)
&&=2\times\frac{e^2}{64\pi (s-m_N^2)^2}\times\bigg(\frac {f_V}{M_V}\bigg)^2\times {\mathbb V}_{h\gamma V}\times 2\kappa^2\times\left[A(K,\kappa_T)+\eta^2 D(K,\kappa_T,\kappa_S)\right]^2\times \tilde{F}(s)\times\frac{8m_N^2}{m_N^2}\,,\nonumber\\
&&=\mathcal{N}^2\times\frac{e^2}{64\pi (s-m_N^2)^2}\times\frac{\left[A(K,\kappa_T)+\eta^2 D(K,\kappa_T,\kappa_S)\right]^2}{A^2(0)}\times \tilde{F}(s)\times 8\,,
\ee
\end{widetext}
with all vertex insertions shown explicitly and, in the last line, we have defined the normalization factor $\mathcal{N}$ as
\bea
\label{diff1N}
\mathcal{N}^2\times e^{2} \equiv 2\times \bigg(\frac {f_V}{M_V}\bigg)^2\times{\mathbb V}_{h\gamma V}\times 2\kappa^2\times A^2(0) \nonumber\\
\eea
for $V=J/\psi$, and
${\mathbb V}_{h\gamma V}=\frac{1}{4\tilde{\kappa}_{V}^4}$ as defined in (\ref{GlueballMesonCoupling}). We have also defined the kinematic factor $$\tilde{F}(s)=(q\cdot p_1)^4\approx \frac{1}{16}\times s^4\,,$$ for $s\gg-t,m_N^2$, a typical signature of a bulk graviton (or spin-2 glueball) exchange.

\begin{figure}[!htb]
\includegraphics[height=5.5cm]{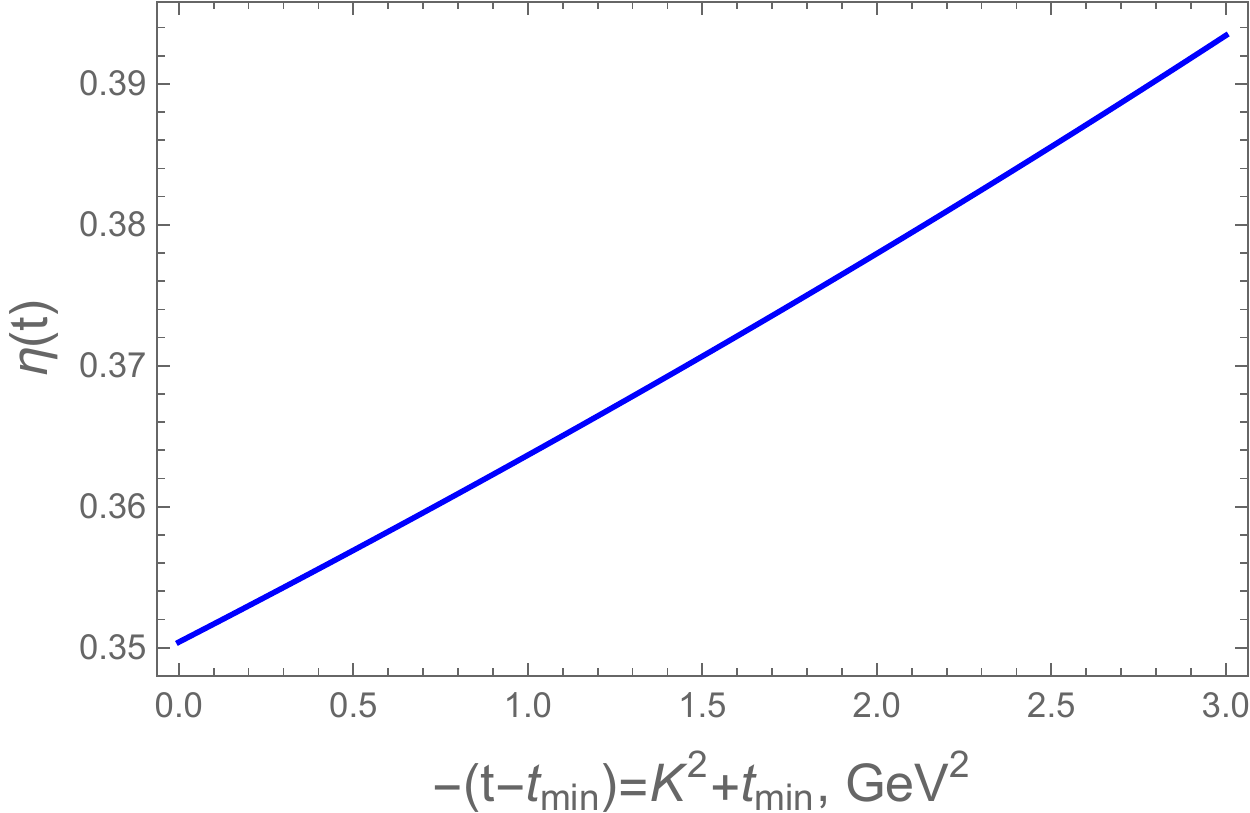}
  \caption{The skewness parameter $\eta$ (defined in (\ref{Eta})) for photoproduction of $J/\Psi$ at $E_\gamma=10.72$ GeV (and with $t_{min}=-0.438~\rm GeV$).}
  \label{fig_eta}
\end{figure}

\begin{figure*}
\subfloat[\label{fig_diff1}]{%
  \includegraphics[height=5.5cm,width=.46\linewidth]{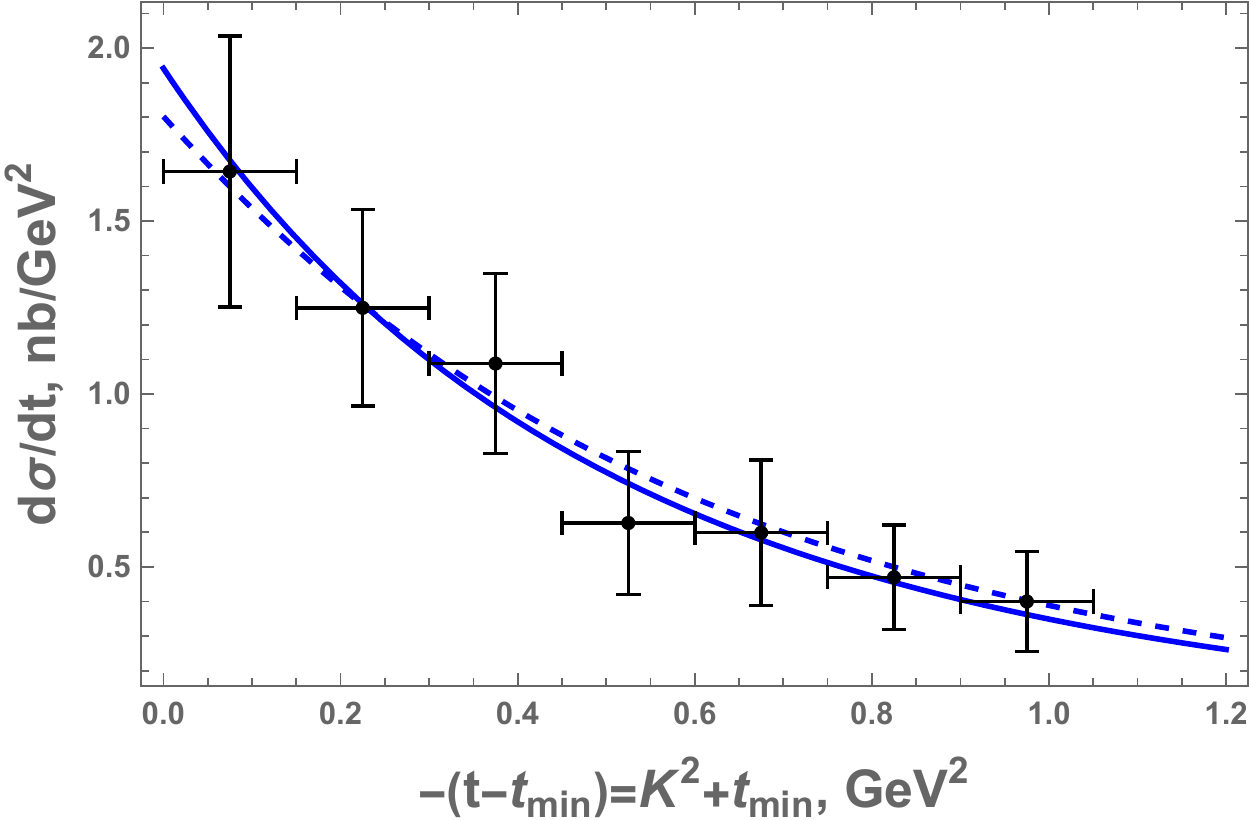}%
}\hfill
\subfloat[\label{fig_diff2}]{%
  \includegraphics[height=5.5cm,width=.46\linewidth]{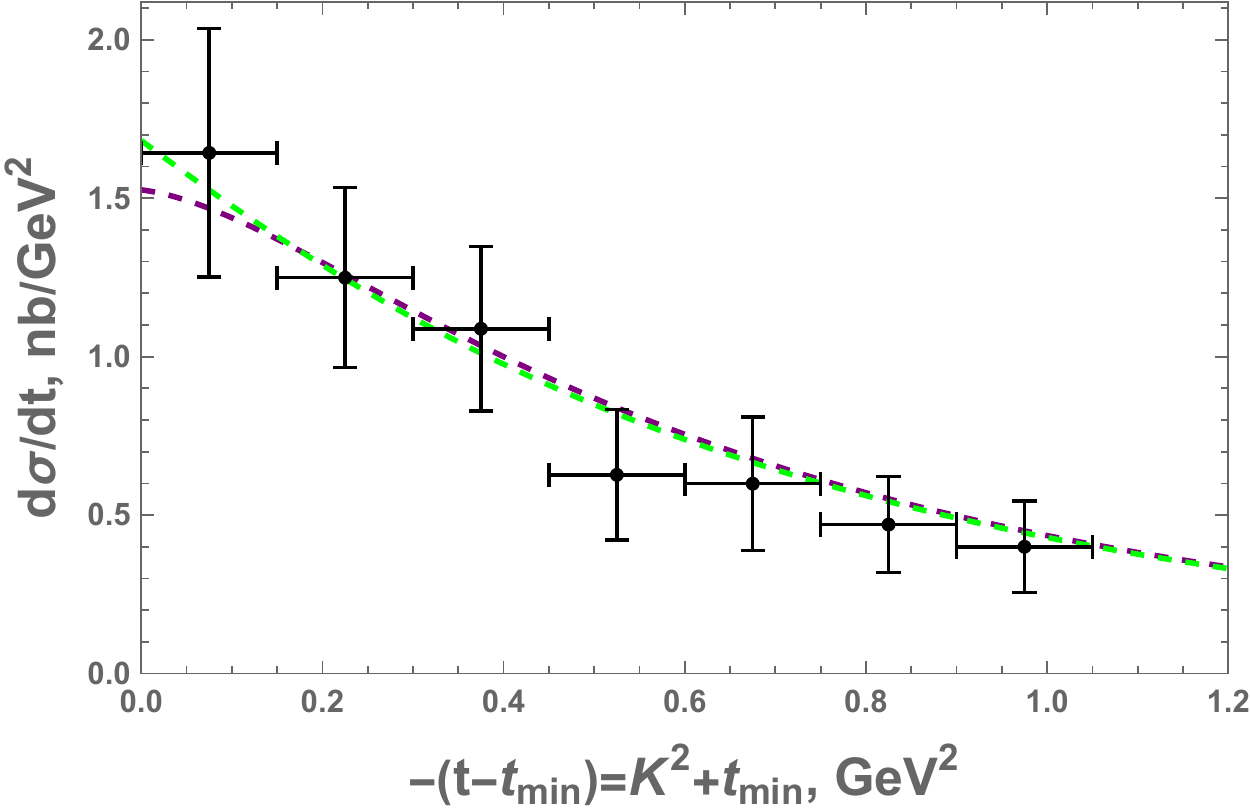}%
}
\caption{(a) Differential cross section  for $V=J/\Psi$ photoproduction for $E_\gamma=10.72$ GeV (and with $t_{min}=-0.438~\rm GeV$). The solid-blue curve is our soft-wall holographic QCD result (\ref{diff1}) with $\mathcal{N}\times e=2.311 \,\text{nb}\,{\rm GeV^{-2}}$, $\kappa_T=\kappa_S=0.388~\text{GeV}$, and $D(0)=-0.231\times A(0)\times m_N^2\times\left({1}/{\kappa_S^2}-{1}/{\kappa_T^2}\right)=0$. The dashed-blue curve is our soft-wall holographic QCD result (\ref{diff1}) with $\mathcal{N}\times e=2.032\,\text{nb}\,{\rm GeV^{-2}}$, $\kappa_T=0.388~\text{GeV}$, $\kappa_S=0.217~\text{GeV}$, and $D(0)=-0.231\times A(0)\times m_N^2\times\left({1}/{\kappa_S^2}-{1}/{\kappa_T^2}\right)=-1.275$. The data is from GlueX~\cite{GlueX:2019mkq}. (b) The dashed-green curve is (\ref{diff1}) with the tripole fits (\ref{Atripole}) to the recent lattice A and D form factors data~\cite{Pefkou:2021fni}. The dashed-purple curve is (\ref{diff1}) with the dipole fits (\ref{Adipole}) to the previous lattice A and D form factors data~\cite{Shanahan:2018pib}. See also Table \ref{table11}.}
  \label{fig_diff12}
\end{figure*}

We note that the kinematic factor (the skewness parameter) $\eta$ is
\begin{equation}
\label{Eta}
\eta =
\frac{M_{V}^2}{4p_1\cdot q - M_{V}^2 + t}\,,
\end{equation}
where $2p_1\cdot q=s-m_N^2$, see Fig.~\ref{fig_eta}. In  general, $\eta\neq \xi$, with  $\xi$ is given by
(see Eq.5.12 in \cite{Belitsky:2005qn} for a very general definition of $\eta$ and $\xi$)
\begin{equation}
\label{Xi}
\xi =
\frac{-M_{V}^2+t/2}{4p_1\cdot q - M_{V}^2 + t}\, .
\end{equation}
In Fig.~\ref{fig_eta}, we show the behavior of the skewness parameter $\eta(t)$ versus $-(t-t_{\rm min})$ at  the photon-point.
From (\ref{diff1}) we conclude that the A-contribution dwarfs the D-contribution by about 10 to 1.

\subsection{Comparison to GlueX}

In Fig.~\ref{fig_diff12}a we show the holographic differential cross section for phtoproduction of $J/\psi$ (\ref{diff1}) versus 
$-(t-t_{\rm min})$ near threshold for a photon energy $E_\gamma=10.72$ GeV. The solid-blue curve is the result 
for $\kappa_T=\kappa_S$ = 0.338 GeV for which the D form factor is null or $D=0$. The  overall normalization
factor (\ref{diff1}) is fixed to  $\mathcal{N}\times e=2.311 \,\text{nb}\,{\rm GeV^{-2}}$. The dashed-blue curve 
is the result for $\kappa_T=0.388~\text{GeV}$, $\kappa_S=0.217~\text{GeV}$, and 
$$D(0)=-0.231\times A(0)\times m_N^2\times\left({1}/{\kappa_S^2}-{1}/{\kappa_T^2}\right)=-1.275\,.$$ The overall normalization factor is now adjusted to $\mathcal{N}\times e=2.032 \,\text{nb}\,{\rm GeV^{-2}}$.
The data are from GlueX~\cite{GlueX:2019mkq}.

For comparison, we show in  Fig.~\ref{fig_diff12}b  the same differential cross section  (\ref{diff1}), but now using the 
lattice A and D gravitational form factors~\cite{Pefkou:2021fni} instead. The dashed-green curve follows from the
lattice A and D using the tripole fits in (\ref{Atripole}), while the dashed-purple curve follows from the lattice 
A  and D using the dipole fits in (\ref{Adipole}).

\section{Conclusion}~\label{CC}

The holographic construction using the soft wall model offers the most economical way of describing the gravitational form
factors of the nucleon as a Dirac fermion in bulk. In the double limit of large $N_c$ and large $^\prime$t Hooft coupling,  the
A form factor follows from the exchange of a bulk-to-boundary graviton, which is dual  to the exchange of Reggeized and
massive $2^{++}$ glueballs at the boundary. Similarly, the D form factor follows from the difference between the exchange
of a bulk-to-boundary graviton and dilaton, dual to Reggeized and massive $2^{++}$ and $0^{++}$ tensor and scalar glueballs
respectively. The holographic B form factor is null.

The holographic gravitational form factor depends on the dilaton profile parameter  $\kappa_T=m_\rho/2=0.388$ GeV, which is fixed by the
rho meson trajectory in bulk (so it is not really a parameter), and the value of 
$A(0)=1+{\cal O}(1/N_c)$ by Poincare symmetry. In the strict large $N_c$ limit,  the D form factor  vanishes 
since the Reggeized $2^{++}$ and $0^{++}$ trajectories are degenerate. At finite $N_c$ they are not,  and the D form factor 
follows by  fixing  $\kappa_S=0.217$ GeV to reproduce the newly released lattice QCD results. Remarkably, holography predicts the
D form factor to be small, negative and tripole-like, while the A form-factor to be large and tripole-like.

By refining the parameter choice of $A(0)$ to the newly released lattice data, we have also sharpened our
holographic predictions for the mass and scalar radii of the proton. A better  estimate of the D form factor 
yields also a sharper description of the pressure and shear across the nucleon. The comparison to the
lattice estimates are in qualitative agreement, although better lattice results  for the D form factor at low
momentum,  would be welcome.

The current GlueX data appears to support the holographic construction in three ways:
1/ the threshold photo-production of $J/\psi$ is dominated by the exchange of a tensor glueball,
as captured by the A form factor; 2/ the contribution of the D form factor appears to be
very  small,  as it is suppressed  by the squared skewness (about $\frac 1{10}$) and its inherent
 smallness (it is null in the strict large $N_c$ limit); The holographic B form factor is null.

Holography provides a 
nonperturbative calculational  framework  based on $1/N_c$ and $1/g^2N_c$ counting, which captures  the stringy character of the gluon interactions both in the
conformal and confined regimes. The QCD string is a reality, that dominates most gluon exchanges in the IR. Detailed  lattice simulations show this string to be surprisingly  close to the Nambu-Goto string. The gravitational excitations in bulk at the origin of the A and D form factors space-like, are the massless modes of this string in a warped space, which allows the interpolation between the conformal UV limit and the confining IR limit. The time-like photoproduction of charmonia and bottomia appears to be dominated by the closed string spin-2 radial Regge trajectory, as captured by the bulk graviton.

Holography is the most economical way of enforcing QCD symmetries, duality and 
crossing symmetries among others, all usually sought by dispersive analysis, and yet within a well defined 
and minimal organisational principle using a stringy field theory.
The present analysis and results are an illustration of that.

However,  holography provides
much more. It allows for the assessment of any n-point function on the boundary,  using 
field theoretical methods through dual Witten diagrams.
It is the QCD string made user friendly.
It also provides for novel physics for processes at low parton-x~\cite{Mamo:2019mka}, and DIS
on nuclei as dual to black holes~\cite{Mamo:2018ync,Mamo:2019jia}.

\vskip 1cm
{\bf Acknowledgements}

We thank Zein-Eddine Meziani for  discussions. K.M. is supported by the U.S.~Department of Energy, Office of Science, Office of Nuclear Physics, contract no.~DE-AC02-06CH11357, and an LDRD initiative at Argonne National Laboratory under Project~No.~2020-0020. I.Z. is supported by the Office of Science, U.S. Department of Energy under Contract No. DE-FG-88ER40388.

\appendix

\begin{widetext}

\section{Details of the electroproduction amplitude}~\label{APP_EE}


\subsection{Wave functions and propagators of bulk virtual photons and vector mesons}

The transverse and longitudinal bulk-to-boundary U(1) propagators $K_{T,L}$ in (\ref{Amph1}) are given by
\bea
K_{T}^{\mu\nu}(q,q^{\prime},\epsilon_T,\epsilon^{\prime}_T,z) &\approx& \epsilon_T\cdot\epsilon_T^{\prime}\times q^\mu q^{\nu}\times\mathcal{V}_{\gamma^{*}}(Q,z)\mathcal{V}_{V}(M_{V},z)
= -q^\mu q^{\nu}\times\mathcal{V}_{\gamma^{*}}(Q,z)\mathcal{V}_{V}(M_{V},z)\,\nonumber\\
&\equiv& q^\mu q^{\nu}\times K_{T}(Q,M_V,z) 
\,,\nonumber\\
K_{L}^{\mu\nu}(q,q^{\prime},\epsilon_L,\epsilon^{\prime}_L,z) &\approx& - \epsilon_L^{\mu}\epsilon_L^{\prime\nu}\times \partial_z\mathcal{V}_{\gamma^{*}}(Q,z)\partial_z\mathcal{V}_{V}(M_V,z) = - q^\mu q^{\nu}\times\frac{1}{Q\,M_V}\times \partial_z\mathcal{V}_{\gamma^{*}}(Q,z)\partial_z\mathcal{V}_{V}(M_V,z)\,\nonumber\\&\equiv& q^\mu q^{\nu}\times K_{L}(Q,M_V,z) \,,\nonumber\\
\label{BKhighenergy}
\eea
where we have defined 
\bea
 K_{T}(Q,M_V,z)&\equiv& -\mathcal{V}_{\gamma^{*}}(Q,z)\mathcal{V}_{V}(M_{V},z)\,\nonumber\\
K_{L}(Q,M_V,z) &\equiv& -\frac{1}{Q\,M_V}\times \partial_z\mathcal{V}_{\gamma^{*}}(Q,z)\partial_z\mathcal{V}_{V}(M_V,z)
\eea
The bulk-to-boundary propagator of the incoming virtual photon $\mathcal{V}_{\gamma^{*}}(Q,z)$, and the normalized wave function of the outgoing $V=J/\psi$ denoted as $\mathcal{V}_{V}(M_V,z)$ are given by
\bea
\mathcal{V}_{\gamma^{*}}(Q,z)
&=&\tilde\xi \,\Gamma(1+a_Q)\,\,{\cal U} (1+a_Q; 2 ; \tilde\xi)
=\tilde\xi\int_{0}^{1}\frac{dx}{(1-x)^2}x^{a_Q}{\rm exp}\Big(-\frac{x\,\tilde\xi}{1-x}\Big)\,,\nonumber\\
\mathcal{V}_{V}(M_V,z)&\equiv&\phi_0(z)=g_5\times\frac{f_V}{M_V}\times 2\tilde\xi L_0^1(\tilde\xi)=g_5\times\frac{f_V}{M_V}\times 2\tilde\xi\,,
\label{vps2swQMV}
\eea
with $a_Q=Q^2/(4\tilde \kappa^2)$ and $\tilde\xi=\tilde{\kappa}_V^2z^2$. The kinematical analysis of the holographic photoproduction amplitude is simplified using $s\gg -t$ 
even at threshold (which is satisfied even for charmonium),  and specifically $p_1^{\mu}\approx p_2^{\mu}$, $q^{\mu}\approx q^{\prime\mu}$, $\epsilon_T\cdot\epsilon_T^{\prime}\approx -1$, $\epsilon_L^{\mu}\approx \frac{q^\mu}{Q}$, $\epsilon_L^{\prime\mu}\approx \frac{q^{\prime\mu}}{M_{V}}$, $\epsilon_{T}\cdot A=\epsilon'_{T}\cdot A\approx 0$ for $A=q,q',p,p_1,p_2$.

\subsection{Spin-2 glueball wavefunctions and propagators}~\label{APP_GB}

In the soft-wall holographic QCD, the normalized wave function for spin-2 glueballs is given by~\cite{Mamo:2019mka}
\bea
 \label{wfSW}
&&J_{h}(m_n,z)\equiv \psi_{n}(z)=c_n\,z^{4}L_{n}^{2}(2\tilde{\xi}_T)\,,\nonumber\\
\eea
with $\tilde{\xi}_T=\kappa_{T}^2z^2$, and normalization
\be
c_n=\Bigg(\frac{2^{4}\kappa_{T}^{6}\Gamma(n+1)}{\Gamma(n+3)}\Bigg)^{\frac 12}=\frac{4\kappa_{T}^{3}}{\sqrt{(n+2)(n+1)}}\,,
\ee
which is determined from the normalization condition (for soft-wall model with background dilaton $\phi=\kappa_{T}^2z^2$)
\be
\int dz\,\sqrt{g}e^{-\phi}\,\abs{g^{xx}}\,\psi_n(z)\psi_m(z)=\delta_{nm}\,.\nonumber\\
\ee
Therefore, defining
\be
F_n=\frac{1}{\sqrt{2}\kappa}\bigg(-\frac{1}{z^{\prime 3}}\partial_{z^\prime}\psi_n(z^\prime)\bigg)_{z^\prime=\epsilon}=-\frac{4}{\sqrt{2}\kappa}c_n L_{n}^{2}(0)\,,\nonumber\\
\ee
with $\psi_n(z\rightarrow 0)\approx c_n\,z^{4}L_{n}^{2}(0)=c_n\,z^{4}\binom{2 +n}{n}$, we can re-write the normalized wave function of glueballs (\ref{wfSW}) in terms of $F_n$ as
\be\label{psigluon}
\psi_{n}(z)=-\sqrt{2}\kappa\times F_{n}\times\frac{1}{4}\times\frac{1}{\kappa_{T}^4}\times\frac{1}{L_{n}^{2}(0)}\times\tilde{\xi}^2 L_{n}^{2}(2\tilde{\xi})\,,
\ee
with $L_{n}^{2}(0)=\binom{2+n}{n}$. Also note that $m_n^2=8\kappa_{T}^2(n+2)$,
\be\label{FN2}
F_n^2=\frac{16}{2\kappa^2}\times \frac{16\kappa_{T}^6}{(n+2)(n+1)}\times \left(L_{n}^{2}(0)\right)^2\,.
\ee

For space-like momenta ($t=k^2=-K^2$), we also have the bulk-to-bulk propagator near the boundary~\cite{Mamo:2019mka}
\be
G(z\rightarrow 0,z^{\prime})\approx \frac{z^4}{4}\sum_n \frac{\sqrt{2}\kappa F_n\psi_n(z^{\prime})}{K^2+m_n^2}=\frac{z^4}{4}\mathcal{H}(K,z^{\prime}) , \nonumber\\\label{hbbt3SW2}
\ee
where, for the soft-wall model~\cite{Mamo:2019mka}

\bea
\mathcal{H}(K,z)&=&\sum_n \frac{\sqrt{2}\kappa F_n\psi_n(z^{\prime})}{K^2+m_n^2}\,,\nonumber\\
&=&4z^{4}\Gamma(\frac{a_K}{2} +2)U\Big(\frac{a_K}{2}+2,3;2\tilde{\xi}_T\Big)
=\Gamma(\frac{a_K}{2}+2)U\Big(\frac{a_K}{2},-1;2\tilde{\xi}_T\Big)\nonumber\\
&=&\frac{\Gamma(\frac{a_K}{2}+2)}{\Gamma(\frac{a_K}{2})}
\int_{0}^{1}dx\,x^{\frac{a_K}{2}-1}(1-x){\rm exp}\Big(-\frac{x}{1-x}(2\tilde{\xi}_T)\Big)\,,
\label{BBSWj2}
\eea
with $\tilde{\xi}_T=\kappa_{T}^2z^2$, $a_K={K^2}/{4\kappa_{T}^2}$, and we have used the transformation $U(m,n;y)=y^{1-n}U(1+m-n,2-n,y)$. (\ref{BBSWj2}) satisfies the normalization condition ${\cal H}(0,z)={\cal H}(K,0)=1$.


The full bulk-to-bulk graviton propagator $G_{\mu\nu\alpha\beta}(m_n,k,z,z^{\prime})$ is given by \cite{Raju:2011mp,DHoker:1999bve}
\be
 \label{Gh}
&&G_{\mu\nu\alpha\beta}(m_n,k,z,z^{\prime})=J_{h}(m_n,z)\times \tilde{G}_{\mu\nu\alpha\beta}(m_n,k)\times J_{h}(m_n,z^{\prime})\,,\nonumber\\
\ee
where the massive spin-2 boundary propagator $\tilde{G}_{\mu\nu\alpha\beta}(m_n,k)$ is given by
\be\label{spin2boundary}
&&\tilde{G}_{\mu\nu\alpha\beta}(m_n,k)=
P_{\mu\nu;\alpha\beta}(m_n,k)\times\frac{-i} {k^2-m_n^2+i\epsilon}\,,
\ee
with the massive spin-2 projection operator $P_{\mu\nu;\alpha\beta}(m_n,k)$ defined as
\be
P_{\mu\nu;\alpha\beta}(m_n,k)={1 \over 2} \left(P_{\mu;\alpha} P_{\nu;\beta} + P_{\mu;\beta} P_{\nu;\alpha} -
\frac 23 P_{\mu;\nu} P_{\alpha;\beta}\right)\,,
\ee
which is written in terms of the massive spin-1 projection operator
\be
P_{\mu;\alpha}(m_n,k)= -\eta_{\mu\alpha} + k_{\mu} k_{\alpha}/m_n^2\,.
\ee

For $z\rightarrow 0$, $t=-K^2$, and focusing on the transverse part of the boundary propagator, i.e., $$\tilde{G}_{\mu\nu\alpha\beta}(m_n,k)\approx\frac{1}{2}\eta_{\mu\alpha}\eta_{\nu\beta}\times \frac{-i} {k^2-m_n^2+i\epsilon}$$ (as we did in~\cite{Mamo:2019mka}), we can simplify (\ref{Amph0}) as

\be
\label{nAmph}
&&i{\cal A}^{h}_{\gamma^{*}_{T/L}p\rightarrow  Vp} (s,t)\approx\frac{1}{g_5}\times(-i)\mathcal{V}^{\mu\nu}_{h\gamma^{*}_{T/L}V}(q,q^{\prime},k)\times \bigg(\frac{-i}{2}\eta_{\mu\alpha}\eta_{\nu\beta}\bigg)\times(-i)\mathcal{V}^{\alpha\beta}_{h\bar\Psi\Psi}(p_1,p_2,k)\,,\nonumber\\
\ee
with
\be
&&\mathcal{V}^{\mu\nu}_{h\gamma^{*}_{T/L}V}(q,q^{\prime},k_)=\sqrt{2\kappa^2}\times\frac{1}{2}\int dz\sqrt{g}\,e^{-\phi}z^4K_{T/L}^{\mu\nu}(q,q^{\prime},\epsilon,\epsilon^{\prime},z)\times\frac{z^4}{4}\,,\nonumber\\
&&\mathcal{V}^{\alpha\beta}_{h\bar\Psi\Psi}(p_1,p_2,k_z)=-\sqrt{2\kappa^2}\times\frac{1}{2}\int dz\,\sqrt{g}\,e^{-\phi}z\,\bar\Psi(p_2,z)\gamma^\mu p^\nu\,\Psi(p_1,z)\times\mathcal{H}(K,z)\,.
\ee


However, if we instead use the full massive spin-2 boundary propagator $\tilde{G}_{\mu\nu\alpha\beta}(m_n,k)$ (\ref{spin2boundary}), we find 
\bea
\label{nAmphfull}
{\cal A}^{h}_{\gamma_{T}^{*}p\rightarrow Vp} (s,t) &=& \frac{1}{g_5}\times 2\kappa^2\times\mathcal{V}_{h\gamma_T^{*}V}(Q,M_V)\times\frac{1}{m_N}\times\bar u(p_2)u(p_1)\nonumber\\
&\times &\sum_{n=0}^{\infty}\,\left[q^{\mu}q^{\nu}\,P_{\mu\nu;\alpha\beta}(m_n,k) \,p_1^{\alpha}p_1^{\beta}\right]\times\mathcal{V}_{h\bar\Psi\Psi}(K,m_n)\,,\nonumber\\
{\cal A}^{h}_{\gamma_{L}^{*}p\rightarrow Vp} (s,t) &=& \frac{1}{g_5}\times 2\kappa^2\times\frac{1}{QM_V}\times\mathcal{V}_{h\gamma_L^{*}V}(Q,M_V)\times\frac{1}{m_N}\times\bar u(p_2)u(p_1)\nonumber\\
&\times &\sum_{n=0}^{\infty}\,\left[q^{\mu}q^{\nu}\,P_{\mu\nu;\alpha\beta}(m_n,k) \,p_1^{\alpha}p_1^{\beta}\right]\times\mathcal{V}_{h\bar\Psi\Psi}(K,m_n)\,,\nonumber\\
\eea
with
\bea
&&\mathcal{V}_{h\gamma^{*}_TV}(Q,M_V)=-\frac{1}{2}\int dz\sqrt{g}\,e^{-\phi}z^4\mathcal{V}_{\gamma^{*}}(Q,z)\mathcal{V}_{V}(M_V,z)\times\frac{z^4}{4}\,,\label{couplingT}\\
&&\mathcal{V}_{h\gamma^{*}_LV}(Q,M_V)=-\frac{1}{2}\int dz\sqrt{g}\,e^{-\phi}z^4\partial_z\mathcal{V}_{\gamma^{*}}(Q,z)\partial_z\mathcal{V}_{V}(M_V,z)\times\frac{z^4}{4}\,,\label{couplingL}\\
&&\mathcal{V}_{h\bar\Psi\Psi}(K,m_n)=-\frac{1}{2}\int dz\,\sqrt{g}\,e^{-\phi}z\,\left(\psi_R^2(z)+\psi_L^2(z)\right)\times\frac{\sqrt{2}\kappa F_{n}\psi_n(z)}{K^2+m_n^2}\,.\label{FFs}\\
\eea

Following~\cite{Nishio:2014eua}, we can evaluate  
\bea\label{spin2contraction}
q^{\mu}q^{\nu}\,P_{\mu\nu;\alpha\beta}(m_n,k) \,p_1^{\alpha}p_1^{\beta}=\left(p_1\cdot q\right)^{2}\times \left(1-\frac{4}{3}\frac{p^2}{m_{n}^2}\times\eta^2\right)\,,
\eea
using the general result (see Eq.279 in~\cite{Nishio:2014eua})
\be\label{spinjcontraction}
q^{\mu_1}q^{\mu_2}...q^{\mu_j}\,P_{\mu_1\mu_2...\mu_j;\nu_1\nu_2...\nu_j}(m_n,k) \,p_1^{\nu_1}p_1^{\nu_2}...p_1^{\nu_j}=\left(p_1\cdot q\right)^{j}\times \hat{d}_{j}(\eta,m_n^2)
\ee
where $\hat{d}_{j}(\eta,m_n^2)$, for even $j=2,4,...$, is a polynomial of skewness $\eta$ of degree $j$ which can be written explicitly in terms of the hypergeometric function $_2F_1(a,b,c,d)$ as 
\be\label{etapoly}
\hat{d}_{j}(\eta,m_n^2)=\, _2F_1\left(-\frac{j}{2},\frac{1-j}{2};\frac{1}{2}-j;-\frac{4p_1^2}{m_n^2}\times\eta ^2\right)\,.
\ee
Note that we have also replaced $p^2\sim p_1^2$ by $-p_1^2$, and $-\Delta^2$ by $m_n^2$ in Eq.279 of~\cite{Nishio:2014eua}, since we are using the massive spin-1 projection operator $P_{\mu;\alpha}(m_n,k)= -\eta_{\mu\alpha} + k_{\mu} k_{\alpha}/m_n^2$ instead of the massless one $P_{\mu;\alpha}(\Delta=k)= \eta_{\mu\alpha} - k_{\mu} k_{\alpha}/\Delta^2$ used in~\cite{Mamo:2019mka} for the conformal case with $\eta_{\mu\alpha}=(-,+,+,+)$ signature. With this in mind and using (\ref{spin2contraction}) in (\ref{nAmphfull}), we obtain  (\ref{nAmphfull2}).

\end{widetext}

\bibliography{radius}


\end{document}